# THE ALPHA CONSTANT FROM RELATIVISTIC GROUPS.

**Gustavo González-Martín**

Departamento de Física, Universidad Simón Bolívar,

Apartado 89000, Caracas 1080-A, Venezuela

Web page at http:\\prof.usb.ve\ggonzalm\

The value of the alpha constant, known to be equal to an algebraic expression in terms of $\pi$ and entire numbers related to certain group volumes, is derived from the relativitistic structure group of a geometric unified theory, its subgroups and corresponding quotients.

# 1. Introduction.

It is known that the fine structure constant $\alpha$ is essentially equal to an algebraic expression in terms of $\pi$ and entire numbers that arises from the quotient of the volume of certain groups [1,2]. This expression may also be obtained, using different physical arguments, starting from the structure group of a unified geometric theory [3]. Using the geometric fact, in this theory, that the tangent space to space time is the image of a Minkowsian subspace of the geometric algebra, the invariant measure in the symmetric space, defined by the associated groups is transported to space time.

# 2. A Geometric Measure.

The current $*J$ is a 3-form on $M$ valued in the Clifford algebra $A$. It is constructed starting from a vector field on the symmetric space $K$. This space is $G/G_+$ where $G$ is the simple group whose action produces the automorphisms of $A$ and $G_+$ is the even subgroup, relative to the orthonormal base of the algebra. The vector field is the image, under the Clifford injection $\kappa$ of a vector field in space time $M$. This injection allows us to define $*J$ as the pullback form of a 3-form in $K$. The integration of the current in a three dimensional boundary of a region $R$ in $M$ is equivalent to the integration of this 3-form pulled back from a geometric form in a three dimensional boundary of the image of the region in the symmetric space $G/G_+$. The latter form is defined by the existence of a geometric invariant measure in $G/G_+$. The constant coefficient of this invariant measure may be calculated in the particular case where the fiber bundle is flat and the field equation reduces to the linear equation equivalent to electromagnetism. This relation defines a geometric interpretation for the coupling constant of the geometric unified theory: "The coupling constant is the constant coefficient of $*J$ introduced by the invariant measure in the symmetric space $G/G_+$".

## 2.1. Symmetric Space *K*.

As indicated before, the group $G$ is SL(4,R) and the even subgroup $G_+$ is $SL_1(2,C)$. The symmetric space $K$ is a non compact real form of the complex symmetric space corresponding to the complex extension of the non compact SU(2,2) and its quotients. The corresponding series of symmetric spaces coincides with the series characterized by the group SO(4,2) as shown in appendix B. In particular we can identify the quotients with the same character, +4, in order to write the series of spaces in the following form,

$$R \equiv \frac{SO(4,2)}{SO(4) \times SO(2)} \approx \frac{SL(4,R)}{SL(2,C) \times SO(2)} \cong K \cong \cdots \approx \frac{SO(6)}{SO(4) \times SO(2)} \quad . \tag{2.1}$$

These quotients include the non compact Riemannian Hermitian $R$ and the non compact pseudo-Riemannian non Hermitian $K$ symmetric spaces. Since some of these groups and quotients are non compact we shall use the normalized invariant measure $\mu_N$ calculated from a known measure, as usually done when working with non compact groups. For compact groups the integral of the invariant measure over the group parameter space gives the group volume. In general, the normalized measure gives only the functional structure of the volume element, in other words, the invariant measure up to a multiplicative constant.

The center of $G$, which is not discrete, contains a generating element $\kappa_5$ whose square is -1. We shall designate by $2J$ the restriction of ad($\kappa_5$) to the tangent space $TK_k$. This space, that has for base the 8 matrices $\kappa_\alpha$, $\kappa_\beta \kappa_5$, is the proper subspace corresponding to the eigenvalue -1 of the operator $J^2$, or,

$$J^2\left(x^\lambda \kappa_\lambda + y^\lambda \kappa_\lambda \kappa_5\right) = \tfrac{1}{4}\left[\kappa_5, \left[\kappa_5, x^\lambda \kappa_\lambda + y^\lambda \kappa_\lambda \kappa_5\right]\right] =$$
$$- x^\lambda \kappa_\lambda - y^\lambda \kappa_\lambda \kappa_5 \quad . \tag{2.2}$$

The endomorphism $J$ defines an almost complex structure over $K$. In addition, using the Killing metric, in the Clifford representation, the complex structure preserves the pseudo Riemannian (Minkowskian) metric. Furthermore the torsion $S$ vanishes,

$$S(a,b) = [a,b] + J[Ja,b] + J[a, Jb] - [Ja, Jb] = 0 \ . \tag{2.3}$$

In this form, the conditions for $J$ to be an integrable complex structure, invariant by $G$ are met and the space $K$ is a non Hermitian complex symmetric space.

## 2.2. Realization of $K$ as a unit polydisc $D^4(K)$.

The bilinear complex metric in $K$ is invariant under SO(4,C) and does not have a definite signature. Using Weyl's unitary trick on the Minkowskian coordinates $x^\lambda$, $y^\lambda$ of the symmetric space $K$, its complex structure is related to the complex structure of $R$. The generators of the quotient $K$ are 2 compact and 6 non compact instead of the 8 non compact generators of the quotient $R$. Both quotients have the matricial structure,

$$K = \begin{bmatrix} \begin{bmatrix} & * & \\ & & \end{bmatrix} & \begin{bmatrix} x^0 & y^0 \\ x^1 & y^1 \\ x^2 & y^2 \\ x^3 & y^3 \end{bmatrix} \\ \begin{bmatrix} & * & \\ & & \end{bmatrix} & \begin{bmatrix} x^4 & y^4 \\ x^5 & y^5 \end{bmatrix} \end{bmatrix} . \tag{2.4}$$

where the lower right submatrix is

$$\begin{bmatrix} x^4 & y^4 \\ x^5 & y^5 \end{bmatrix} = \begin{bmatrix} 1 + x \bullet x & x \bullet y \\ y \bullet x & 1 + y \bullet y \end{bmatrix}^{\frac{1}{2}} . \tag{2.5}$$

The conditions imposed by the associated groups SL(4,R) and SO(4,2) over the corresponding coordinates on these spaces, expressed by the scalar product in this submatrix, are related respectively by the Minkowskian and Euclidian metrics.

Define the six complex coordinates $t^a$ on $R$ that relate this space to the complex space $C^6$, where $R$ is inmersed, that is,

$$t^a = x^a + i y^a \quad 0 \leq a \leq 5 \ . \tag{2.6}$$

These coordinates may be expressed in terms of the four corresponding coordinates $u^\alpha$ on $K$ by recognizing the scalar product in eq. (2.5),

$$\delta_{\mu\nu} t^\mu t^\nu \leftrightarrow -\eta_{\mu\nu} u^\mu u^\nu = I \kappa_{(\mu} \kappa_{\nu)} \quad 0 \leq \mu \leq 3 \ , \tag{2.7}$$

and if we introduce new coordinates $t$ on $K$,

$$t^m = u^m \ , \tag{2.8}$$

$$t^0 = i u^0 \ , \tag{2.9}$$

we find the same conditions on the coordinates $t$ on $K$ that exist on the coordinates $t$ on $R$.

The conditions over the coordinates $t^4$ and $t^5$ allow us to reduce to $C^5$ the complex space where the realization of $K$ is immersed. If we introduce the four complex projective coordinates $z^\mu$, we obtain the realization,

$$z^\mu = \frac{t^\mu}{t^4 - it^5} \qquad 0 \leq \mu \leq 3 \quad . \tag{2.10}$$

If we indicate the transposed by $z'$ the conditions on these coordinates are those of the unit polydisc, $D^4 \subset D^5$, defined by

$$D^n(K) = \{ z \in C^n ; \; 1 + |zz'|^2 - 2\bar{z}z' > 0, \; |zz'| < 1 \} \quad . \tag{2.11}$$

In this manner, the complex coordinates define a holomorphic diffeomorphism $h$ of $K$ onto the interior of a bounded symmetric domain $D$. The bounded realization of the space $K$ is the unit polydisc $D^4(K)$. This realization $D^4(K)$ corresponds to the bounded realization of the space $R$, the unit polydisc $D^4(R)$ by a change in coordinates. Although the interior of $D^4$ is not compact we can apply the existing mathematical techniques of the classical bounded domains to study the space $K$, in particular we can find normalized invariant measures for the spaces $K$ and $R$.

## 2.3. Invariant Measure on the Polydisc.

A geometric measure on the space $K$, an 8 dimensional hyperboloid $H^8$, arises from a measure on $C^5$ in a manner similar as the measure on the Euclidian spheres is obtained from a measure on $R^n$. In order to evaluate this measure, it is convenient to use the immersion, $i : D^4 \to D^5$ defined on the intersection of the $D^5$ and the plane $z^5=0$. Since $D^5$ is a homogeneous space under the action of the compact group $SO(5) \times SO(2)$, using this group and $SO(5,2)$ we may obtain the measure on the quotient $R$, [4], which is equal the measure on $K$.

In order to construct these measures it is convenient to define certain domains related to $D^5$, [5]. Silov's boundary, the generalization of the circle as the boundary of the 1 dimensional complex disk, is established by the Fourier transformation on the symmetric space $D^n$. It is the characteristic space of $D^n$, in other words it allows us to characterize the holomorphic functions on $D^n$ by their value on this boundary. It is defined by

$$Q^n(K) = \{ \xi = xe^{i\theta}; \; x \in R^n, \; |xx'| = 1, \; 0 \leq \theta \leq \pi \} \quad . \tag{2.12}$$

Poisson's kernel $P_n(z, \xi)$ over $D^n \times Q^n$ is defined as the Euclidian invariant measure on the characteristic space $Q^n$. This kernel has the value

$$P_n(z, \xi) = \frac{\left(1 + |zz'|^2 - 2\bar{z}z'\right)^{n/2}}{V(Q^n) \times \left|(z - \xi)(z - \xi)'\right|^n} \quad , \tag{2.13}$$

determined by Hua [5]. The actual construction of the measure over $Q^4$, due to Wyler, is indicated in the appendix A. The expression for the harmonic functions over $D^n$ is

$$\varphi(z) = \int_{Q^n} P_n(z, \xi) \varphi(\xi) d\xi \quad , \tag{2.14}$$

which, for the case of the disc, reduces to a solution the Dirichlet problem using Poisson's integral formula that gives the harmonic function knowing its value on circle boundary,

$$\varphi(z) = \int_0^{2\pi} P(z, \theta) \varphi(\theta) d\theta \quad . \tag{2.15}$$

The Poisson kernel defines a normalized form $\mu_N$ over the characteristic space because

$$\int_{Q^n} P_n(z,\xi)d\xi = 1 \quad . \tag{2.16}$$

As indicated above, the measure of interest over the polydisc $D^4$, representative of the hyperboloid $K$, is obtained from the complex space $C^5$ where it is immersed. There are injective mappings,

$$M^4 \xrightarrow{\kappa} K^8 \xrightarrow{h} D^4 \xrightarrow{i} D^5 \quad , \tag{2.17}$$

that correspond to Clifford's mapping $\kappa$ and the holomorphic mapping $h$. The immersion $i:D^4 \rightarrow D^5$ allows us to pull back the Euclidian measure on the characteristic space $Q^5$, boundary of $D^5$, to $Q^4$ and then to corresponding boundaries in $K$ and $M$. The form $\mu_N$ on the image of $M$, defines, in this way, a geometric form on space time,

$$(h \circ \kappa)^* \mu_N [M^4] = \mu_N [h_* \kappa_* M^4] \quad . \tag{2.18}$$

## 3. Value of the Geometric Coefficient.

We may associate a physical current 3-form $^*J$ to the standard unnormalized volume form $\mu$. Similarly, in a natural form, we may associate another geometric current 3-form $^*J_g$ to the normalized geometric form $\mu_{Ng}$ with the same factor or constant coefficient $\alpha_g$. In addition, since the physical current form should not be associated to a normalized volume form, this geometric current 3-form should be defined by additionally multiplying by the volume of a characteristic space determined by the physical solutions. Let us define $^*J_g$ on the boundary of a region $R$ in $M$,

$$\begin{aligned} J_g &\equiv V(\partial R) \times (h \circ \kappa)^* \mu_{Ng}[\partial R] = V(\partial R)\alpha_g \times (h \circ \kappa)^* \mu[\partial R] \\ &= V(\partial R)\alpha_g \times J \end{aligned} \quad . \tag{3.1}$$

The constant coefficient in this equation may be identified, once and for all, using any solution. If we assume staticity conditions and spherical symmetry that allow a decomposition of space time $M^4$ in two orthogonal subspaces, spatial spheres $S^2$ and the supplementary space time $M^2$, the forms decompose in two components and it becomes easier to calculate the constant coefficient,

$$(h \circ \kappa)^* \mu_{Ng}[M^4] = \kappa^* h^* \mu_{Ng}[M^2] \wedge \kappa^* h^* \mu_{Ng}[S^2] \quad . \tag{3.2}$$

The geometric charge $Q_g$, given by the induced geometric measure, is the geometric calculation of interest. It represents physically the integral of the geometric current $\alpha j$, obtained by absorbing the constant $\alpha$ in the current, over a spatial hypersurface $\sigma \supset M$ that contains the $S^2$ subspace. Let us restrict the problem to the special particular case of pure electromagnetism in a flat space time $M$.

All solutions of this restricted problem may be found as a sum of fundamental solutions that correspond to the Green's function for the electromagnetic field. The Green's function determines the field of a point source which always corresponds to a spherically symmetric static field relative to an observer at rest with the source. Thus the restricted problem reduces to Coulomb's problem in flat three dimensional space. This is precisely the situation where the $\alpha$ constant, or equivalently Coulomb's constant, is introduced. The spherically symmetric harmonic potential solutions are determined, using Poisson's integral formula, by its value on a boundary sphere. We see that the characteristic boundary space, where integration is performed to find solutions of the restricted physical problem, is the sphere $S^2$ in $R^3$. Geometrically, in accordance with our theory, the form $h^*\mu_{Ng}$ should be restricted to its component over the image of a sphere, determined by the Clifford mapping $\kappa$, in the space $K$. Therefore, the characteristic space is $\kappa(S^2)$. We have then the volume of this characteristic space,

$$V(\kappa(S^2)) = \int_{\kappa S^2} h^* \mu_{Ng} = 2\int_{S^2} \kappa^* h^* \mu_{Ng} = 8\pi \quad . \tag{3.3}$$

The volume of $\kappa(S^2)$, indicated in the previous equation, is twice the volume of $S^2$ due to the 2-1 homomorphism between standard spinors and vectors determined by its homomorphic groups SU(2) and SO(3). Hence the geometric coefficient $\alpha$ is

$$\alpha = \alpha_g V(\kappa_* S^2) = 8\pi\alpha_g \quad . \tag{3.4}$$

The value of this constant coefficient is obtained substituting, in the last equation, the value of Wyler's coefficient calculated in appendix A,

$$\alpha[M^4] = \frac{2^3 \pi 3^2}{2^6 \pi^5}\left(\frac{\pi^5}{2^4 \times 5!}\right)^{\frac{1}{4}} = \frac{9}{16\pi^3}\left(\frac{\pi}{120}\right)^{\frac{1}{4}} = \frac{1}{137.03608245} = \alpha_p \quad , \tag{3.5}$$

which is equal to the experimental physical value of the alpha constant. The field equation may be written geometrically,

$$D^*\Omega = 4\pi\alpha^* J = 4\pi^* J_g \quad . \tag{3.6}$$

# 4. Conclusions.

The coupling constant of the geometric unified theory may be calculated from the volumes of certain symmetric spaces related to the structure group of the theory and its subgroups. There is no additional arbitrary constant to be used in the theory.

# Appendix A.

In what follows, we indicate Wyler's calculation of the value of the constant coefficient of the measure on $Q^4$, Silov's boundary of $D^4$. This measure is obtained by constructing Poisson's measure invariant under general coordinate transformations in $D^5$ by the group of analytic mappings of $D^5$ onto itself which is SO(5,2).

The calculation is based on the following proposition: The isotropy subgroup at the origin, SO(5)×SO(2), acts transitively over $Q^5$ and Poisson's kernel $P_n(z,\xi)$, harmonic on $D^n$, represents an invariant measure of the action of SO(5)×SO(2) over $Q^5$.

Although $P_4(x,\xi)$ represents a measure $\mu$ on $Q^4$, it is not adequate in this case because we need the measure defined by the Euclidean measure in one dimension higher, $C^5$, the one induced from $Q^5$. Since $P_5(z,\xi)$ is an invariant Euclidean measure over $Q^5$ we construct the induced measure by the immersion i:$Q^4 \rightarrow Q^5$ which is equal to the kernel $P_4(z,\xi)$ up to constant coefficient factor. Functionally both measures are equivalent.

The invariant normalized form over the characteristic space is

$$\mu_N[Q^5] = P_5(z,\xi)d\xi = \frac{\Pi_5(z,\xi)}{V(Q^5)} = \frac{\mu[Q^5]}{V(Q^5)} \quad , \tag{A.1}$$

$$\int_{Q^5} \mu_N = \int_{Q^5} P_5(z,\xi)d\xi = 1 \quad , \tag{A.2}$$

where $\mu$ represents an invariant form, not normalized, over $Q^5$ defined by these equations. The action of SO(5,2) over the Hermitian structure of $D^5$ defines the Bergman metric on this space. The group SO(5,2), of coordinate transformations, acts on $D^5$ and consequently on $Q^5$. The measure defined by the Poisson kernel is not invariant under SO(5,2), but is related to the invariant metric measure by

$$\mu_{Ng}[Q^5] = P_5(z,\xi)j_g d^5\xi = \frac{j_g}{V(Q^5)}\mu[Q^5] \quad , \tag{A.3}$$

in terms of $j_g$, which denotes the complex Jacobian or determinant of the Jacobian matrix $JC$ of the mapping $z \to G(z)$ where $G$=SO(5,2).

To find $j_g$ we use a relation, given by Hua, between the Bergman kernel and the volume density over the domain $D^5$. The Bergman kernel may be written as

$$B_5 = \frac{1}{V(D^5) \times (1 + |zz'|^2 - 2\bar{z}z')^n} = \frac{|\det J_C(z)|^2}{V(D^5)} \quad . \tag{A.4}$$

The real Bergman metric $h$ is defined by the invariant bilinear form

$$ds^2 = \frac{d\bar{z}\bar{J}'_C(z)J_C(z)dz'}{V(D^5)} \quad . \tag{A.5}$$

Thus, the value of the complex Jacobian of the transformation $z \to G(z)$ is

$$j_g = (\det J_C) = (\det J_R)^{\frac{1}{2}} = (\det h)^{-\frac{1}{4}} = (V(D^5))^{\frac{1}{4}} \quad , \tag{A.6}$$

obtaining by substitution in eq. (A.3) the metric measure,

$$\mu_{Ng}[Q^5] = P_5(z,\xi)j_g d^5\xi = \frac{(V(D^5))^{\frac{1}{4}}}{V(Q^5)}\mu[Q^5] \quad . \tag{A.7}$$

To obtain the Wyler measure, in $Q^4$, it is necessary to reduce the action of the isotropy group I(5,2) to the isotropy subgroup I(4,2),

$$\mu_{Ng}[Q^4] = \frac{(V(D^5))^{\frac{1}{4}}}{V(Q^5)} \frac{V[I(4,2)]}{V[I(5,2)]}\mu[Q^4] = \alpha_g[Q^4] \times \mu[Q^4] \quad . \tag{A.8}$$

The inverse of the measure of the isotropy groups quotient is

$$\frac{V[SO(5) \times SO(2)]}{V[SO(4) \times SO(2)]} = \frac{V[SO(5)]}{V[SO(4)]} = V(S^4) = \frac{2\pi^{5/2}}{\Gamma(5/2)} = \frac{2^3 \pi^2}{3} \quad . \tag{A.9}$$

Under this reduction the coefficient of Poisson's kernel over $Q^4$, the constant coefficient of the normalized measure $\mu_N$ in eq. (A.8), defines the coefficient of the measure over $D^4$,

$$\alpha_g[Q^4] = \frac{(V(D^5))^{\frac{1}{4}}}{V(Q^5)} \frac{V[I(4,2)]}{V[I(5,2)]} = \frac{(V(D^5))^{\frac{1}{4}} \times V(SO(4))}{V(Q^5) \times V(SO(5))} \quad . \tag{A.10}$$

The indicated volumes are known. The volume of the polydisc is

$$(D^n) = \frac{\pi^n}{2^{n-1} n!} \tag{A.11}$$

and the volume of Silov's boundary is the inverse of the coefficient in the Poisson kernel,

$$(Q^n) = \frac{2\pi^{\frac{n}{2}+1}}{\Gamma(\frac{n}{2})} \quad . \tag{A.12}$$

In particular we have,

$$(D^5) = \frac{\pi^5}{2^4 \times 5!} \quad , \tag{A.13}$$

$$(Q^5) = \frac{2^3 \pi^3}{3} \quad . \tag{A.14}$$

Substitution of theses expressions in equation (A.10) gives Wyler's coefficient of the induced invariant measure,

$$\alpha_g[Q^4] = \frac{3^2}{2^6 \pi^5} \left( \frac{\pi^5}{2^4 \times 5!} \right)^{\frac{1}{4}} \quad . \tag{A.15}$$

## Appendix B

Here we indicate the relationship between the spaces $R$ and $K$, as components of a series of symmetric spaces characterized by the group SO(4,2).

The Cartan Killing metric for a group quotient space, $G'/H$, is taken as the metric in the subspace of the algebra of $G'$, complementary to the algebra of $H$. The exponentiation of this subspace is a globally symmetric space [4] because any point and its neighborhood can be translated to any other point by a group operation. In this way it is possible to show that the metric is invariant.

Since both the group $G'$ and the subgroup $H$ are related to compact groups by means of involutive automorphisms, there are different quotients related among each other by two involutions that we shall indicate as $\sigma$ and $\tau$. The possibilities for both involutions are exactly the same available for the involutive automorphisms of the complex extension of the algebra. Hence, the classification of the symmetric spaces is determined applying a pair of these involutions, taken from the indicated set, that commute and exhaust the possibilities. The simultaneous eigenvalues of this pair serve to describe the Lie algebra $A$,

$$A = A_{++} \oplus A_{+-} \oplus A_{-+} \oplus A_{--} \quad . \tag{B.3}$$

The involutive automorphism $\tau$ serves to select a compact subgroup, starting from a compact group $G$

$$G_{+\tau} = \exp(A_{++} \oplus A_{-+}) \tag{B.4}$$

and the compact symmetric space with definite metric

$$\exp(i(A_{+-} \oplus A_{--})) = \frac{G}{G_{+\tau}} \quad . \tag{B.5}$$

The other automorphism $\sigma$ serves to convert the compact subgroup to a non compact subgroup

$$G_{+t} = \exp(A_{++} \oplus A_{-+}) \xrightarrow{\sigma} \exp(A_{++} \oplus iA_{-+}) = G_{+\tau}^{\sigma} \tag{B.6}$$

and to convert the symmetric space with definite metric in one with indefinite metric

$$\frac{G}{G_{+\tau}} = \exp(i(A_{++} \oplus A_{--})) \xrightarrow{\sigma} \exp(i(A_{+-} \oplus iA_{--})) = \frac{G^{\sigma}}{G_{+\tau}^{\sigma}} \quad . \tag{B.7}$$

The symmetric space $G/G_+$ has a negative definite metric derived from the Cartan Killing metric. The dual space $G^*/G_+$, where $G_+$ is the maximal compact subgroup, has an equal but positive definite metric, derived from the Cartan Killing metric restricted to the complementary subspace of $A^*$. Both spaces are, therefore, Riemannian spaces.

There is a theorem that says that the non compact irreducible Hermitian symmetric spaces are exactly the manifolds $G/H$ where $G$ is a connected non compact simple group with center $\{I\}$ and $H$ is a maximal compact subgroup of $G$ with a non discrete center [6]. There is a standard notation for the classification of Riemannian spaces, indicating the Cartan subspace (A,B…), the involution type (I,II,III) and the dimensions that characterize the groups (2n,p,q).

There are other real forms of the complex group $G^C$ between the two extremes $G$ y $G^*$ and therefore there is a series of symmetric spaces that are real forms of the complex extension of the quotient,

$$\left(G/G_+\right)^C = G^C/G_+^C \quad , \tag{B.8}$$

that fall between the two extreme Riemannian spaces. These intermediate spaces have an indefinite metric and are considered pseudo Riemannian spaces. The different real forms within the series corresponding to a complex symmetric space are classified by their characters. The character of a real form is defined as the trace of the canonical form of the metric. This integer, corresponds to the difference in the number of compact and non compact generators. The series may be characterized by its non compact end group.

The series of quotient spaces related to the $A_3$ Cartan space are of interest. In particular we choose the involutive automorphism $\tau$ of type AIII(p=2,q=2) that determines a seven dimensional compact subgroup $G_+$. We obtain, in this manner, a series of eight dimensional spaces, characterized by the non compact group SU(2,2), corresponding to the Riemannian space $G/G_+$ and its dual $G^*/G_+$,

$$\frac{SU(4)}{SU(2) \otimes SU(2) \otimes U(1)} \approx \frac{SU^*(4)}{SL(2,C) \otimes SO(2)} \approx \frac{SU(2,2)}{SL(2,C) \otimes SO(1,1)} \approx$$
$$\approx \frac{SL(4,R)}{SL(2,C) \otimes SO(2)} \approx \frac{SU(2,2)}{SU(2) \otimes SU(2) \otimes U(1)} \quad . \tag{B.9}$$

Due to the isomorphism of the spaces $A_3$ and $D_3$ we have the isomorphic series, characterized by the non compact group SO(4,2), corresponding to the Riemannian coset space $G/G_+$ and its dual $G^*/G_+$, with involution $\tau$ of the type BDI(p=4,q=2),

$$\frac{SO(6)}{SO(4) \otimes SO(2)} \approx \frac{SO(5,1)}{SO(3,1) \otimes SO(2)} \approx \frac{SO(4,2)}{SO(3,1) \otimes SO(1,1)}$$

$$\approx \frac{SO(3,3)}{SO(3,1) \otimes SO(2)} \approx \frac{SO(4,2)}{SO(4) \otimes SO(2)} \quad . \tag{B.10}$$

The characters of the real forms of both isomorphic series are -8, -4, 0, +4, +8.